\documentclass[a4paper,12pt,numbers,sort&compress]{article}

\pdfoutput=1

\usepackage[paper=letterpaper,margin=1in]{geometry}

\usepackage{amsmath,amssymb,amsfonts,epsfig,cite,setspace,bigstrut,longtable,array,breqn,url,color}
\usepackage{todonotes}

\usepackage{hyperref,caption,setspace}

\usepackage{pdfpages,lipsum}

\begin{document}



\vskip 0.25in

\newcommand{\sref}[1]{\S~\ref{#1}}
\newcommand{\nn}{\nonumber}
\newcommand{\tr}{\mathop{\rm Tr}}
\newcommand{\comment}[1]{}

\newcommand{\cM}{{\cal M}}
\newcommand{\cW}{{\cal W}}
\newcommand{\cN}{{\cal N}}
\newcommand{\cH}{{\cal H}}
\newcommand{\cK}{{\cal K}}
\newcommand{\cY}{{\cal Y}}
\newcommand{\cZ}{{\cal Z}}
\newcommand{\cO}{{\cal O}}
\newcommand{\cA}{{\cal A}}
\newcommand{\cB}{{\cal B}}
\newcommand{\cC}{{\cal C}}
\newcommand{\cD}{{\cal D}}
\newcommand{\cE}{{\cal E}}
\newcommand{\cF}{{\cal F}}
\newcommand{\cX}{{\cal X}}
\newcommand{\IA}{\mathbb{A}}
\newcommand{\IP}{\mathbb{P}}
\newcommand{\IQ}{\mathbb{Q}}
\newcommand{\IH}{\mathbb{H}}
\newcommand{\IR}{\mathbb{R}}
\newcommand{\IC}{\mathbb{C}}
\newcommand{\IF}{\mathbb{F}}
\newcommand{\IV}{\mathbb{V}}
\newcommand{\II}{\mathbb{I}}
\newcommand{\IZ}{\mathbb{Z}}
\newcommand{\re}{{\rm~Re}}
\newcommand{\im}{{\rm~Im}}

\newcommand{\tmat}[1]{{\tiny \left(\begin{matrix} #1 \end{matrix}\right)}}
\newcommand{\mat}[1]{\left(\begin{matrix} #1 \end{matrix}\right)}

\let\oldthebibliography=\thebibliography
\let\endoldthebibliography=\endthebibliography
\renewenvironment{thebibliography}[1]{%
\begin{oldthebibliography}{#1}%
\setlength{\parskip}{0ex}%
\setlength{\itemsep}{0ex}%
}%
{%
\end{oldthebibliography}%
}

\newtheorem{theorem}{\bf THEOREM}
\def\thetheorem{\thesection.\arabic{theorem}}
\newtheorem{proposition}{\bf PROPOSITION}
\def\thetheorem{\thesection.\arabic{proposition}}
\newtheorem{observation}{\bf OBSERVATION}
\def\thetheorem{\thesection.\arabic{observation}}

\def\theequation{\thesection.\arabic{equation}}
\newcommand{\setall}{\setcounter{equation}{0}
        \setcounter{theorem}{0}}
\newcommand{\setequation}{\setcounter{equation}{0}}
\renewcommand{\thefootnote}{\fnsymbol{footnote}}

~\\
\vskip 1cm

\begin{center}
{\Large \bf  Deep-Learning the Landscape}

\medskip
\vspace{4mm}

{\large Yang-Hui He}

\vspace{1mm}

\renewcommand{\arraystretch}{0.5} 
{\small
{\it
\begin{tabular}{rl}
  ${}^{1}$ &
  Department of Mathematics, City, University of London, EC1V 0HB, UK\\
  ${}^{2}$ &
  Merton College, University of Oxford, OX14JD, UK\\
  ${}^{3}$ &
  School of Physics, NanKai University, Tianjin, 300071, P.R.~China\\
\end{tabular}
}
~\\
~\\
~\\
hey@maths.ox.ac.uk
}
\renewcommand{\arraystretch}{1.5} 

\end{center}

\vspace{10mm}

\begin{abstract}
We propose a paradigm to deep-learn the ever-expanding databases which have emerged in mathematical physics and particle phenomenology, 
as diverse as the statistics of string vacua or combinatorial and algebraic geometry.
As concrete examples, we establish multi-layer neural networks as both classifiers and predictors and train them with a host of available data
ranging from  Calabi-Yau manifolds and vector bundles, to quiver representations for gauge theories.
We find that even a relatively simple neural network can learn many significant quantities to astounding accuracy in a matter of minutes and can also predict hithertofore unencountered results.
This paradigm should prove a valuable tool in various investigations in landscapes in physics as well as pure mathematics.

\end{abstract}


\newpage

\tableofcontents

\newpage

\section{Introduction and Summary}
Theoretical physics now firmly resides within an Age wherein new physics, new mathematics and new data coexist in a symbiosis which transcends inter-disciplinary boundaries and wherein concepts and developments in one field are evermore rapidly enriching another.
String theory has spearheaded this  vision for the past few decades and has, perhaps consequently, become a paragon of the theoretical sciences.
That she engenders the cross-fertilization between physics and mathematics is without dispute: interactions on an unprecedented scale have commingled fields as diverse as quantum field theory, general relativity, condensed matter physics, algebraic and differential geometry, number theory, representation theory, category theory, etc.
With the advent of increasingly powerful computers, from this fruitful dialogue has also arisen a plethora of data, ripe for mathematical experimentation.

This emergence of data in some sense began with the incipience of string phenomenology \cite{Candelas:1985en} where compactification of the heterotic string on Calabi-Yau threefolds (CY3) was widely believed to hold the ultimate geometric unification.
A race, spanning the 1990s, to explicitly construct examples of Calabi-Yau (CY) manifolds ensued, beginning \footnote{
The interested readers should find the opportunity to have Prof.~Philip Candelas FRS, relate to them the fascinating account of how this was achieved on the then CERN super-computer, involving dot-matrix printers and magnetic tapes， or, Prof.~Rolf Schimmrigk, on staring at VAX machines in Austin over long nights. 
} with the so-called complete intersection CY manifolds (CICYs) \cite{cicy}, proceeding to the hypersurfaces in weighted projective space \cite{wp4}, to elliptic fibrations \cite{ellip} and ultimately culminating in the impressive (at least some $10^{10}$) list of CY3s from reflexive polytopes \cite{ks}.

With the realization that the landscape of stringy vacua \footnote{
The now-popular figure of some $10^{500}$ is a back-of-envelop estimate based on a generic number of cycles in a generic CY3 that could support fluxes.
} might in fact exceed the number of inequivalent Calabi-Yau threefolds \cite{Kachru:2003aw} by hundreds of orders of magnitude, there was a vering of direction toward a more multi-verse or anthropic philosophy.
Nevertheless, hints have emerged that the vastness of the landscape might well be mostly infertile (cf.~the swamp-land of \cite{Vafa:2005ui}) and that we could live in a very special universe \cite{Gmeiner:2005vz,Braun:2005nv,Candelas:2007ac}, a ``des res'' corner within a barren vista.

Thus, undaunted by the seeming over-abundance of possible vacua, fortified by the rapid growth of computing power and inspirited by the omnipresence of big data, the first two decades of the new millennium saw a return to the earlier principle of creating and mining geometrical data;
the notable fruits of this combined effort between pure and computational algebraic geometers as well as formally and phenomenologically inclined physicists have included  (q.v.~\cite{He:2013epn} for a review of the various databases)
\begin{itemize}
\item Continuing with the Kreuzer-Skarke database, exemplified by
	\begin{itemize}
	\item Finding quotients to reach small Hodge numbers \cite{Candelas:2008wb};
	\item Improvement of PALP, the original KS computer programme \cite{Braun:2012vh} and incorporation into SAGE \cite{sage};
	\item Identifying refined structures \cite{Taylor:2012dr,Candelas:2012uu,He:2015fif};
	\item Interactive Calabi-Yau databases as websites \cite{cyDB,cyDBus};
	\item building line-bundles and monad bundles \cite{He:2009wi} for heterotic phenomenology in generalizing embedding;
	\end{itemize}
\item Generalizing the CICY construction, exemplified by
	\begin{itemize}
	\item Porting the CICY database into Mathematica and establishing a catalogue of stable bundles \cite{Anderson:2007nc};
	\item A database of MSSM and 3-generation models from heterotic compactification \cite{Anderson:2013xka,Anderson:2012yf,Gao:2014nfa};
	\item Relaxing the ambient Fano requirement \cite{Anderson:2015iia};
	\item CICY 4-folds \cite{Gray:2013mja};
	\end{itemize}
\item Finding elliptic and K3 fibred CY for F-theory and string dualities \cite{Morrison:1996na,Friedman:1997yq}, exemplified by
	\begin{itemize}
	\item Identifying elliptic fibrations \cite{Braun:2011ux,Taylor:2012dr} from KS data;
	\item Establishing phenomenologically viable dataset of stable bundles using spectral covers \cite{Gabella:2008id,Gao:2014nfa};
	\item studying finiteness in the elliptic landscape \cite{Cvetic:2014gia};
	\end{itemize}
\item D-brane world-volume theories as supersymmetric quiver gauge theories, exemplified by
	 \begin{itemize}
	\item Classifications of quivers from $\IC^3$ and $\IC^4$ orbifolds \cite{Hanany:1998sd};
	\item non-compact toric CY and brane-tilings \cite{Feng:2000mi,Franco:2005sm} and their databases \cite{Davey:2009bp,Franco:2017jeo,Davey:2011mz,Franco:2016qxh};
	\item specializing the toric CY to the reflexive polytopes, whereby linking with KS data \cite{Hanany:2012hi,He:2017gam}.
	\end{itemize}
\end{itemize}

All of the above cases are accompanied by typically accessible data of considerable size.
To the collection of these databases, representing a concrete glimpse onto the string landscape, we shall refer as {\bf landscape data}, for lack of better words.
For instance, the heterotic line bundles on CICYs are on the order of $10^{10}$, the spectral-cover  bundles on the elliptically fibred Calabi-Yau, $10^6$, the brane-configurations in the Calabi-Yau volume studies, $10^5$, type II intersecting brane models, $10^9$, etc.
Even by today's measure, these constitute a fertile playground of data, the likes of which Google and IBM are constantly analysing.
A natural course of action, therefore, is to do unto this landscape data, what Google et al.~do each second of our lives: to machine-learn.

Let us be precise about what we mean by {\em deep machine-learning}
this landscape.
Much of the aforementioned data have been the brain-child of the marriage between physicists and mathematicians, especially incarnated by applications of computational algebraic geometry, numerical algebraic geometry and combinatorial geometry to problems which arise from the classification in the physics and recast into a finite, algorithmic problem in the mathematics (cf.~\cite{comp-book}).
Obviously, computing power is a crucial limitation.

Unfortunately, in computational algebraic geometry - on which most of the data heavily rely, ranging from bundles stability in heterotic compactification to Hilbert series in brane gauge theories - the decisive step is finding a Groebner basis, which is notoriously known to be unparallelizable and double-exponential in running time.
Thus, much of the challenge in establishing the landscape data had been to either circumvent the direct calculation of the Groebner bases by harnessing of the geometric configuration -- e.g., using the combinatorics when dealing with toric varieties. Still, many of the combinatorial calculations, be they triangulation of polytopes or finding dual cones, are still exponentially expensive.

The good news for our present purpose is that, {\em much of the data have already been collected}.
Oftentimes, as we shall find out in our forthcoming case-studies, tremendous effort is needed for deceptively simple questions.
Hence, to draw inferences from {\it actual} theoretical data by deep-learning therefrom would not only help identify undiscovered patterns but also aid in predicting results which would otherwise cost formidable computations.
Subsequently, we propose our
\begin{quote}
{\bf Paradigm: }
To set-up neural networks to deep-learn the landscape data, to recognize unforeseeable patterns (as classifiers) and to extrapolate to new results (as predictors).
\end{quote}

Of course, this paradigm is useful not only to physicists but to also to mathematicians; for instance, could our neural work be trained well enough to approximate bundle cohomology calculations? This, and a host of other examples, we will shortly examine.
Indeed, such neural networks have recently been applied to data analyses in other fields of physics, such as high energy experiment \cite{Leney:2014dya}, astrophysics/cosmology \cite{Novaes:2014ska}, energy landscapes \cite{dhagash}, and holography \cite{Gan:2017nyt}.
Of note is a recent conference in the role of deep-learning in fundamental physics \cite{conf,conf2}.
It is therefore timely that we should employ this powerful tool to string theory/mathematical physics as well as to computational problems in pure mathematics.

The organization of the paper is as follows.
We begin in \S\ref{s:nn} with a bird's-eye-view of the principles of machine deep-learning and neural networks with emphasis on its implementations by the latest version of Mathematica so that the readers could familiarize themselves with this handy tool for their own entertainment.

Then, in \S\ref{s:eg} we establish various readily adapted and easily constructed neural networks of the multi-layer perceptron type for a host of examples from the above-mentioned databases, ranging from Calabi-Yau manifolds, to vector bundles, to quiver gauge theories, amply demonstrating the usefulness of our paradigm.

Throughout we shall be explicit about the layers and training as well as test data, adhering to Mathematica syntax for concreteness and familiarity, whilst keeping track of computation time and accuracy achieved.
We beg the readers' indulgence in including code, which is perhaps uncustomary in a paper in mathematical physics; however, the nature of our investigations permits little alternative.
We conclude with prospects, including a list of specific directions and issues, in \S\ref{s:conc}.

This paper is a detailed expansion of the brief companion letter \cite{He:2017dia} which summarizes the key results.
Here, we will explain both the data and the methodology in detail, as well as give a pedagogical review of neural networks and machine learning to theoretical and mathematical physicists.

\section{Neural Networks and Deep-Learning}\label{s:nn}
We begin with some rudiments of deep-learning and neural networks; this by no means embodies an attempt to be a pedagogical introduction to this vast field but only serves to set some notations and concepts.
Emphasis will be placed on their implementation in the newest versions of Mathematica \cite{Freeman,math}, rather than on a general theory.
We refer the reader to the classic texts \cite{HKP,Hassoun,Haykin} on the subject.
Indeed, there are standard software available for the {\sf python} programming language which is standard amongst experts, such as {\sf Tensorflow} \cite{tensorflow}, however, since Mathematica is the canonical software for theoretical physicists, we will adhere to tutorial using this familiar playground.

The prototypical problem which an artificial intelligence (AI) might encounter is one of text recognition. For example, how does a machine recognize the individual digits from
$
\includegraphics[trim=0mm 0mm 0mm 0mm, clip, width=3in]{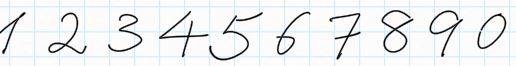}
$
?
One could try to geometrically define the numbers by their homotopy and other invariants, but this is especially difficult given that hand-writing has such huge variation.

In some sense, many problems in string theory and in algebraic geometry are of a similar nature.
For example, to compute the cohomology of a vector bundle on an algebraic variety -- which physically could encode the number of generations of particles in a gauge theory -- is a highly non-trivial task of sequence-chasing, but in the end produces a few non-negative integers, often rather small.

The premise of machine learning is to take an data-driven perspective by sampling over large quantities of hand-written digits and feeding them into an optimization routine, whose basic components are called {\it neurons} in analogy to the human brain.
When there is a large collection - often organized in {\it layers} -  of interconnected neurons, we have a {\bf neural network}; the more the number of layers, or more generally the greater the complexity of the inter-connectivity amongst the neurons, the ``deeper'' the learning.

\subsection{The Perceptron}
The archetypal neuron is the so-called {\it perceptron} (also called the single-layer perceptron (SLP) to emphasize its simplicity);
this is a function $f(z_i)$ (called {\it activation function}) of some input vector $z_i$.
The function $f$ is usually taken to be binary, or some approximation thereof, such as by the hyperbolic tangent $f(z) = \tanh(z)$ or the logistic sigmoid function 
\begin{equation}
\label{sigmoid}
f(z) = \sigma(z) := \left(1 + \exp(-z) \right)^{-1} \ , 
\end{equation}
both of which we shall use liberally later.
The activation function is set to contain real parameters, say of the form $f(\sum\limits_i w_iz_i + b)$, where $w_i$ are called weights and $b$, the bias.

Supose we are given {\it training data} $D = \{ (x_i^{(j)}, d^{(j)} \}$, which is a collection, labelled by $j$, of inputs $x_i^{(j)}$ and {\em known} outputs $d^{(j)}$.
We then consider minimizing on the parameters $w_i$ and $b$, by some method of steepest descent, so that the error (standard deviation) 
\begin{equation}
SD = 
\sum\limits_j \left( f(\sum_i w_i x_i^{(j)} + b) - d^{(j)} \right)^2
\end{equation}
is minimized.
We resort to numerical minimization because of the number of parameters could be very large in general, especially when there are many input channels and/or layers.
Note that at this simple stage, the principle is very similar to a regression model in statistics, with the difference that the weight parameters $w_i$ are updated with the introduction of each subsequent new data: this is the neuron ``learning''.

In Mathematica versions 11.0 and above, the {\sf Neuronetworks} package had been updated and integrated into the system functions;
however, since it is still experimental and much of the documentation is rather skeletal, it behoves us to  briefly describe the implementation (in all subsequently examples, we will adhere to Mathematica notation) as follows: 
\begin{itemize}
\item
We begin by defining the SLP  with \verb| NetChain[ ] |, with argument a list containing a single element (we will generalize this in the next subsection) specifying the activation function in the layer;
\item
  The layer can be a linear layer specifying a linear transformation by a matrix of specified dimension, for example
  \[
  \verb|net = NetChain[{LinearLayer[5]}];|
  \]
  or by an appropriate activation function,
  for instance\\
  \verb|net = NetChain[{ElementwiseLayer[Ramp]}]|, \\ or
  \verb|net = NetChain[{ElementwiseLayer[Tanh]}]|, \\ or
  \verb|net = NetChain[{ElementwiseLayer[LogisticSigmoid]}]|.
\item
We can initialize the SLP, now called \verb|net|, with some random weights and bias via 
\verb|net = NetInitialize[net]|;
\item
The neural network can now be trained with actual data.
Since with a single layer not too much can be expected, we will discuss the training in the ensuing sub-section, after introducing multi-layers.
\end{itemize}

We remark that in the above, the syntax for the linear layer can  simply be abbreviated to \verb|net = NetChain[{5}]|.
Likewise, the word \verb|ElementWiseLayer| can also be dropped, as in
\verb|net = NetChain[{Ramp}]|.
Indeed, many other more sophisticated operations can also be chosen for the layer, such as \verb|ConvulutionLayer[ ]|, \verb|LongShortTermMemoryLayer[ ]|, etc.,
and we refer the reader to \cite{stackex} for some more pedagogical explanations as well as a useful lexicon of layers.

\subsection{Neural Networks}\label{s:digit}
An SLP can be chained together to a multiple layer perceptron (MLP).
Sometimes MLPs and neural networks are used inter-changeably though it should be noted that MLPs are only a special type of neural networks, viz., feed-forward, because it is a uni-directional chain.
In other words, one joins a list of SLPs in a directed fashion to produce an MLP, while an neural work is a directed graph (with potential cycles to allow for feed-back) each node of which is an SLP, analogous to the brain being a complex directed graph of neurons.

Nevertheless, with an MLP, we can already go quite far \footnote{We further remark that there are built-in functionalities in Mathematica such as Classify[~] and Predict[~] which perform similar tasks. However, because they are black-box neural networks, we will adhere to our explicit MLP throughout this paper for illustrative purposes.}.
To return to our hand-writing example, one could resort to many resources on the internet, such as maintained by the Modified National Institute of Standards and Technology (MNIST) database, to which Mathematica can link.
The data is simply a string of substitution rules of the form ``writing sample'' $\to$ ``correct identification'':
\[
\verb|data = |
\{
\includegraphics[trim=0mm 0mm 0mm 0mm, clip, width=3in]{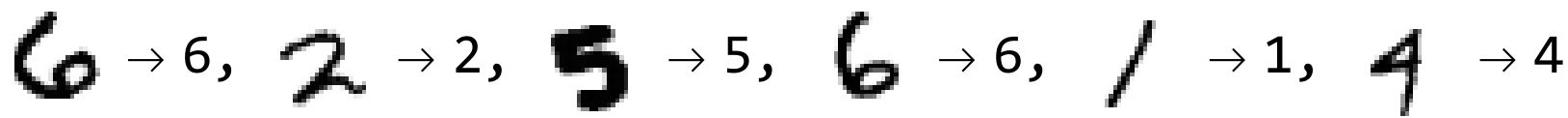}
\ldots 
\}
\]
We now follow Wolfram's documentation for this example.
The procedure, as will be throughout the paper, will be in four steps:
\begin{description}

\item[(1) Data Acquisition: ]
First, the data can be obtained online directly, from which we take a sample of 100 of a total of some 60,000: 
\begin{verbatim}
resource = ResourceObject["MNIST"];
data = ResourceData[resource, "TrainingData"];
sample = RandomSample[data, 100];
\end{verbatim}

\item[(2) Establishing Neural Network: ] 
Next, an MLP can then be set up as
\begin{verbatim}
net = NetChain[{ConvolutionLayer[20, 5], Ramp, PoolingLayer[2, 2],
         ConvolutionLayer[50, 5], Ramp, PoolingLayer[2, 2],
         FlattenLayer[], 500, Ramp, 10, SoftmaxLayer[]}];
\end{verbatim}

\item[(3) Training Neural Network: ]
Finally, training the neural network is simply done with the command \verb|NetTrain[ ]| as
\begin{verbatim}
net = NetTrain[net, sample];
\end{verbatim}
Here is where one optimizes on all the parameters (weights and biases) in each layer, aiming to reduce error (standard deviation) as more data is encountered.
One can check that even learning our relatively small sample gives us rather good results:
\begin{verbatim}
Table[{net[sample[[i, 1]]], sample[[i, 1]]}, {i, 1, 100}]
\end{verbatim}
gives almost all correct answers.

\item[(4) Testing Unseen Data/Validation: ]
A caveat of NNs is that because there are so many parametres over which we optimize that there is the danger of {\bf over-fitting}.
In the above example, we have seen 100 training data points and the NN has fitted to them completely.
It is therefore crucial to let the trained NN test against data which is {\it not} part of the training set, i.e., data {\it unseen} by the NN.

In our example, the training set is of size 100. We now try against a validation set of size, say 1000, as follows,
\begin{verbatim}
validation = RandomSample[data, 1000];
Length[Select[Table[{net[validation[[i, 1]]], validation[[i, 2]]}, 
                             {i, 1, 1000}], #[[1]] == #[[2]]&]]
\end{verbatim}
We find that the result is 808 (this is, of course, a typical value because we have randomized over samples), meaning that the NN has learnt, with a mere 100 data-points, and managed to predict correct values to an accuracy of about 80\%, on 1000 data points.
This is quite impressive indeed!

\end{description}

There is obviously much flexibility in choosing the various layers and some experimentation may be required (q.v.~\cite{stackex}).
What we have described above is usually called {\bf supervised} learning because there is a definite input and output structure; we wish to train the NN so that it could be useful in predicting the output given unencountered input values.
This is obvious usefulness to the problems in physics and mathematics at hand and now we will turn to a variety of case studies.

\section{Case Studies}\label{s:eg}
With the tool presented above, we now proceed to analyse our landscape data, a fertile ground constituting more than 2 decades of many international collaborations between physics and mathematicians.
Our purpose is to first ``learn'' from the inherent structure and then to ``predict'' unseen properties; considering how difficult some of the calculations involved had been in establishing the databases, such quick estimates, even if qualitative, would be highly useful indeed.

\subsection{Warm-up: Calabi-Yau Hypersurfaces in $W\IP^4$}
One of the first datasets \cite{wp4} to experimentally illustrate mirror symmetry was that of hypersurfaces in weighted projective space $W\IP^4$, which, though later subsumed by the Kreuzer-Skarke list \cite{ks} since $W\IP^4$ is toric, is a good starting point because of the simplicity in definition.
Given the ambient space $W\IP^4_{[w_0 : w_1 : w_2 : w_3 : w_4]} \simeq (\IC^5 - \{0\}) / \sim$ where the equivalence relation is 
$( x_0, x_1, x_2, x_3, x_4) \sim (\lambda^{w_0}x_0, \lambda^{w_1}x_1, \lambda^{w_2}x_2, \lambda^{w_3}x_3, \lambda^{w_4}x_4)$ for
points $( x_0, x_1, x_2, x_3, x_4)$ in $\IC^5$, $\lambda \in \IC^*$ and weights
$w_{i=0,\ldots,4} \in \IZ_+$.
This space is in general singular, but choosing a generic enough  homogeneous polynomial of degree $\sum\limits_{i=0}^4 w_i$ which misses the singularities, this polynomial defines a hypersurface therein which is a smooth Calabi-Yau threefold $X$.

\begin{figure}[!h!t!b]
\begin{picture}(500,350)(0,-160)
\put(60,0){$\chi  = 2(h^{1,1} - h^{2,1})$}
\put(80,170){$h^{1,1} + h^{2,1}$}
\put(360,0){$\chi$}
\put(220,120){\#}
%
%
\centerline{
\begin{tabular}{cc}
(a)
\includegraphics[trim=0mm 0mm 0mm 0mm, clip, width=3in]{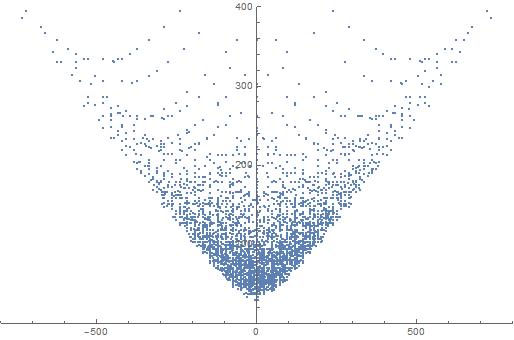}
&
(b)
\includegraphics[trim=0mm 0mm 0mm 0mm, clip, width=3in]{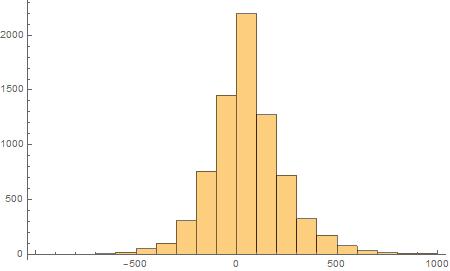}
\\
&\\
&\\
(c)
\includegraphics[trim=0mm 0mm 0mm 0mm, clip, width=3in]{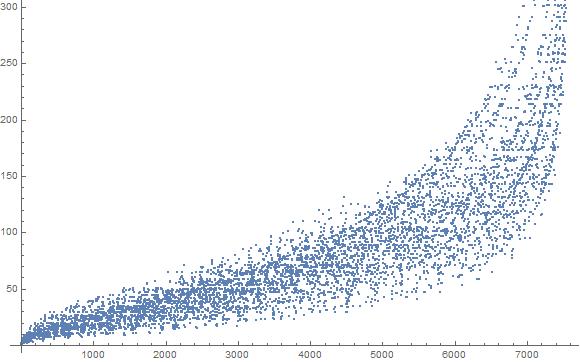}
&
(d)
\includegraphics[trim=0mm 0mm 0mm 0mm, clip, width=3in]{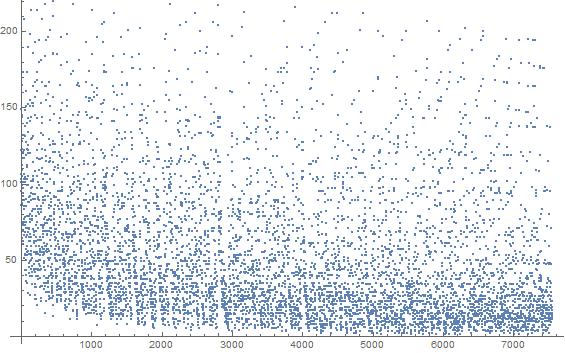}
\end{tabular}
}
\end{picture}
\caption{{\sf {\small From the set of 7555 Calabi-Yau threefold hypersurfaces in $W\IP^4$, we plot
(a) $\chi  = 2(h^{1,1} - h^{2,1})$ against the sum $(h^{1,1} + h^{2,1})$, exhibiting mirror symmetry;
(b) a histogram of $\chi$, showing a concentration of self-mirror manifolds at $\chi=0$;
(c) an ordered list-plot of $h^{1,1}$ and (d) that of $h^{2,1}$. 
}}
\label{f:wp4data}}
\end{figure}

There are 7555 inequivalent such configurations, each specified by a 5-vector $\vec{w}_{i=0,\ldots,4}$.
The Euler characteristic $\chi$ of $X$ is easily given in terms of the vector. However, as is usually the case, the individual Hodge numbers $(h^{1,1}, h^{2,1})$ are less amenable to a simple combinatorial formula (even though $\chi = 2(h^{1,1} - h^{2,1})$ is).
In fact, the original computation had to resort to  Landau-Ginzberg techniques from physics to obtain the full list of Hodge numbers \cite{wp4}.
One could  in principle use a combination of the adjunction formula and Euler sequences, in addition to singularity resolution, but this is not an easy task to automate.

In Figure \ref{f:wp4data}, we present some standard statistics of the data.
In part (a), we plot the traditional $\chi  = 2(h^{1,1} - h^{2,1})$ against the sum $(h^{1,1} + h^{2,1})$, exhibiting mirror symmetry;
the histogram of $\chi$ is drawn in part (b).
In parts (c) and (d), for reference, we draw the $h^{1,1}$ and $h^{2,1}$ values against the order as presented in the database of \cite{cyDB};
this is to illustrate that while $h^{1,1}$ has been organized more or less in an increasing order, $h^{2,1}$ appears rather random.

\subsubsection{Deep-Learning Complex Structure}
Let us try to deep-learn the least suggestive set of data, i.e., $h^{2,1}$, which encodes the complex structure of the CY3, and presented in Figure \ref{f:wp4data}(d); the human eye cannot quite discern what patterns may exist therein.
We let \verb|WPcy3| be the set of training data, being a list of
\[
\{ \mbox{configuration (5-vector)}, \ \chi, \ h^{1,1}, \ h^{2,1} \} \ , 
\]
so that a typical element would be 
something like $\{\{1,3,5,9,13\},-128,14, 78 \}$.
Suppose we have a simple (qualitative) question: {\it how many such CY3s have a relatively large number of complex deformations}?
We can, for instance, consider $h^{2,1} > 50$ to be ``large'' and set up our training data to be
\begin{verbatim}
ddd = Table[WPcy3[[i, 1]]   ->   If[(WPcy3[[i, 4]] > 50, 1, 0],
         {i, 1, Length[WPcy3]}];
\end{verbatim}

Next, we set up a neural network (MLP) with the input being a 5-vector and the output a number, and consisting of 5 total layers, 3 of which are intermediate (called ``hidden layers'' in the literature):
\begin{equation}\label{wp4net}
\begin{array}{l}
\verb|net = NetChain[|\\
\verb|     {LinearLayer[100], ElementwiseLayer[LogisticSigmoid], |\\
\verb|      ElementwiseLayer[Tanh], LinearLayer[100], SummationLayer[]}, |\\
\verb|      "Input" -> {5}];|
\end{array}
\end{equation}
The layers are self-explanatory. The input (first) layer is a linear function, essentially a $5 \times 100$ matrix, followed by a sigmoid function from \eqref{sigmoid}, then by a hyperbolic tangent, both of which are elementary-wise operations, and another linear layer, here an $100 \times 100$ matrix, finishing with a weighted sum of the 100-vector into a number.

We now train, with say 500 iterations, our MLP by the data \verb|ddd| as:
\begin{verbatim}
net = NetTrain[net, ddd, MaxTrainingRounds -> 500];
\end{verbatim}
and in under a minute on an ordinary laptop, the MLP is trained.

To appreciate the accuracy of the training we can compare the predicted result of the network against the actual data.
We do so by counting the number of cases where the net value for a given input {\it differs} from the correct value, for instance, by
\begin{verbatim}
Select[Table[{Round[net[ddd[[i, 1]]]], ddd[[i, 2]]}, {i, 1, 7555}], 
         #[[1]] != #[[2]] &] // Length
\end{verbatim}
Note that while we have set the data to be binary: 1 if $h^{2,1} > 50$ and 0 otherwise, the neural network does not know this and has, especially with the use of the analytic functions in layers 2 and 3, optimized the weights and baises  to give a (continuous) real output. 
Hence, we use round-off to do the comparison.
The answer to the above is 282.
In other words, our neural network has, in under one minute, learnt - from a seemingly random set of data - to answer the question of deciding whether a given  CY3 has a large number of complex parameters, to an accuracy of $(7555 - 282)/7555 \simeq 96.2 \%$. 
This is re-assuring, in the sense that this dataset seems to be ``learnable'' (we will see, in a later section, how certain problems seem intractable).

\paragraph{Learning Curve: }
Having gained some confidence, we can now move to the crux of the problem: {\it how does our NN behave on unseen data?}
This would give one an appreciation of the {\em predictive} power of the network. 
Suppose that we only had partial data.
This is particularly relevant when for instance, due to computational limitations, a classification is not yet complete, or when the quantity in question, here $h^{2,1}$, has not been or could not be yet computed.
Therefore, let us pretend that we have only data available for a random selection of 3000 out of the 7555 $(X, h^{2,1})$ pairs, i.e.,
\begin{verbatim}
net = NetTrain[net, RandomSample[ddd, 3000], MaxTrainingRounds -> 500];
\end{verbatim}
This is less than a half of the total.

Suppose now we are given a new configuration, one which we have not encountered before, with what confidence could we predict that it would have a large number for $h^{2,1}$?
We repeat the last two commands in the above, using \verb|NetTrain[ ]| on the 3000, and then testing against the full 7555.
We find that only 421 cases were actually wrong.
Thus, with rather incomplete training data (about some 40\% of the total number in the classification), the relatively simple neural network consisting of a feed-forward 5 layer MLP, has learnt, in under a minute, our question and predicted new results to 94.4\% accuracy.
One standard measure of the goodness of fit is the so-called {\bf cosine distance}, where we take the normalized dot product between the predicted value, as a vector $v_NN$ (here of length 7555) with the correct values, also as a vector $v_{data}$:
\begin{equation}
d(v_{NN}, v_{data}) := \frac{v_{NN} \cdot v_{data}}{|v_{NN}|\ |v_{data}|} \in [-1,1] \subset \IR \ .
\end{equation} 
Therefore, a value close to $+1$ is a good fit.
In the above example, we find $d = 0.91$.

\begin{figure}[!h!t!b]
\centerline{
\includegraphics[trim=15mm 0mm 0mm 0mm, clip, width=6in]{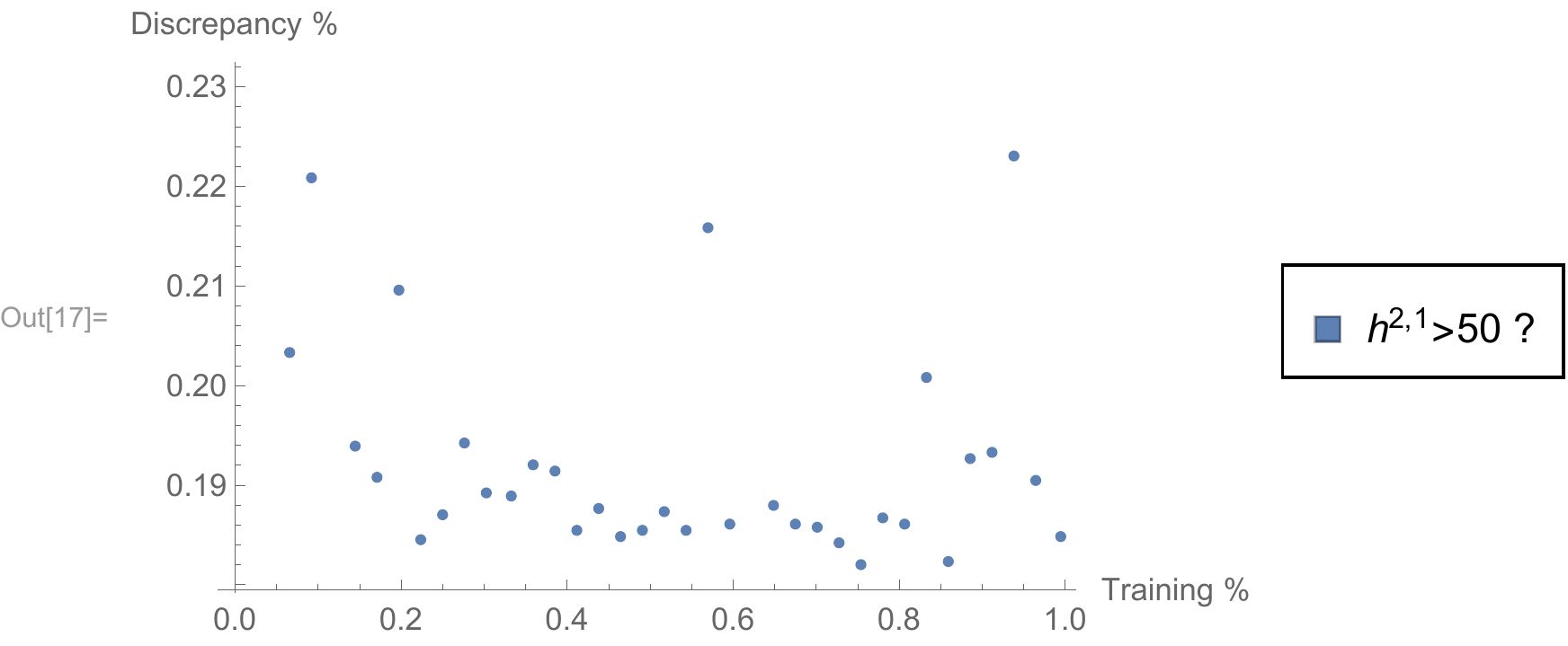}
}
\caption{{\sf {\small 
The learning curve for large complex structure (whether $h^{2,1} > 50$) for the hypersurface Calabi-Yau threefolds in weighted projective $\IP^4$.
We take, in increments of 200, and starting from 500 random samples, with which we train the NN (shown as percentage of the total data size of 7555 in the horizontal).
We then check against the full data to see the percentage discrepancy, shown in the vertical.
}}
\label{f:WPtrainingC}
}
\end{figure}

This partial data paradigm, in splitting the data into (1) training set and (2) validation set, is an important one.
An illustrative quantity is the so-called {\bf learning curve}: we take random samples of increasing size and the check the percentage discrepancy (or cosine distance) of the predicted data versus actual data.
We do this for our example, in intervals of say, size 200, starting from 500 and present this curve in Figure {f:WPtrainingC}.
The axis of abscissa is the \% of total data-size (here 7555) which is used to train the NN and the ordinate is the \% of discrepancy.
We see that even at a low percentage of trained data, we achieved around 90\% accuracy, increasing to about 96\% to the full set.
We will see different shapes of curves later, but this problem is particularly well-suited for prediction since with a relatively small available data one can achieve a good fit.

\subsubsection{Deep-Learning Number of Generations}
Emboldened, let us move onto another question, of importance here to string phenomenology:
{\it Given a configuration, can one tell whether $\chi$ is a multiple of 3?}
In the early days of heterotic string compactifications, using what has now come to be known as the standard embedding, this question was decisive on
whether the model admitted 3 generation of particles in the low-energy effective gauge theory.
Of course, we have an analytic formula for $\chi$ in terms of the input 5-vector $w_i$.
However, testing divisibility is not an immediate task.

Again, we can define a binary function taking the value of 1 if $\chi \bmod 3 \equiv 0$ and 0 otherwise.
Training with the network in \eqref{wp4net}, we achieve 82\% accuracy with 1000 training rounds, taking about 2 minutes; this means that such a divisibility problem is a it less learnable than the previous problem (we will again address and contrast the class of problems later).
With partial data, say taking a random sample of 3000 to train, and validating against the full set, gives also about 80\% correct results.

The astute reader might question at this stage why we have adhered to {\it binary queries}. 
Why not train the network to answer a direct query, i.e., to try for instance 
to learn and predict the value of $h^{2,1}$ itself? 
This is a matter of spread in the present dataset: we have only some $10^4$ inputs but we can see that
the values of $h^{1,1}$ ranges from $1$ to almost $500$; the neural network would have to learn from a relatively small sample in order to distinguish 
some 500 output channels.
We simply do not have enough data here to make more accurate statements, though we will shortly proceed to multiple-channel output.
One can check that net-training on the direct data
\begin{verbatim}
ddd=Table[WPcy3[[i, 1]]  ->  WPcy3[[i, 4]], {i,1,Length[WPcy3]}];
\end{verbatim}
would give the rough trend but not the precise value of the complex structure.

This is precisely in line with our philosophy, the power of deep-learning the landscape lies in rapid {\em estimates}.
The computation of the exact values in the landscape data often takes tremendous efforts in algebraic geometry and quantum field theory, and we wish to learn and benefit from the fruits the labour from the last 2 decades.
We aim to identify patterns and draw inferences and to avoid the intense computations.

\subsection{CICYs}
Having warmed up with the hypersurfaces in weighted projective $W\IP^4$, let us move onto our next case, the CICY dataset, of complete intersection Calabi-Yau threefolds in products of (unweighted) projective spaces.
This is both the first Calabi-Yau database (or, for that matter, the first database in algebraic geometry) \cite{cicy} and the most heavily studied recently for phenomenology \cite{Anderson:2007nc,Anderson:2013xka,Anderson:2015iia,Anderson:2012yf,Gao:2014nfa}.
It has the obvious advantage that the ambient space is smooth by choice and no singularity resolution is needed.
The reason we study this after the $W\IP^4$ data is because, as we see shortly, the input is a configuration of non-negative integer matrices, one rank up in complexity from the 5-vector input in the previous subsection.

Briefly, CICYs embedded as $K$ homogeneous polynomials in $\IP^{n_1} \times \ldots \times \IP^{n_m}$, of multi-degree $q_j^r$.
Here, complete intersection means that the dimension of the ambient space exceeds the number $K$ of defining equations by precisely 3, i.e., $K=\sum\limits_{r=1}^m n_r-3$. Moreover, the Calabi-Yau condition of vanishing first Chern class of $TX$ translates to $\sum\limits_{j=1}^K q^{r}_{j} = n_r + 1 \ \forall \; r=1, \ldots, m$.
Subsequently, each manifold can be written as an $m \times K$ configuration matrix (to which we may sometimes adjoin the first column, designating the ambient product of projective spaces, for clarity; we should bear in mind that this column is redundant because of the CY condition):
\begin{equation}\label{cicy}
X = 
\left[\begin{array}{c|cccc}
  \IP^{n_1} & q_{1}^{1} & q_{2}^{1} & \ldots & q_{K}^{1} \\
  \IP^{n_2} & q_{1}^{2} & q_{2}^{2} & \ldots & q_{K}^{2} \\
  \vdots & \vdots & \vdots & \ddots & \vdots \\
  \IP^{n_m} & q_{1}^{m} & q_{2}^{m} & \ldots & q_{K}^{m} \\
  \end{array}\right]_{m \times K \ ,}
\quad
\begin{array}{l}
q_j^r \in \IZ_{\geq 0} ;  \\
K = \sum\limits_{r=1}^m n_r-3 \ , \\
\sum\limits_{j=1}^K q^{r}_{j} = n_r + 1 \ , \ \forall \; r=1, \ldots, m \ .
\end{array}
\end{equation}
The most famous CICY is, of course, $[4|5]$ or simply the matrix $[5]$, denoting the quintic hypersurface in $\IP^4$. 

The construction of CICYs is thus reduced to a combinatorial problem of classifying the integer matrices in \eqref{cicy}.
It was shown that such configurations are finite in number and the best available computer at the time (1990's), viz., the super-computer at CERN \cite{cicy},
was employed.
A total of 7890 inequivalent manifolds were found, corresponding to matrices with entries $q_j^r \in [0,5]$, of size ranging from $1 \times 1$ to maximum number of rows and columns being 12 and 15, respectively.
In a way, this representation is much closer to our archetypal example of hand-writing recognition in \S\ref{s:digit} than one might imagine.
The standard way to represent an image is to pixelate it, into blocks of $m \times n$, each of which carrying a colour info, for example, a 3-vector encapturing the RGB data.

\begin{figure}[!h!t!b]
\centerline{
\begin{tabular}{cc}
(a)
\includegraphics[trim=0mm 0mm 0mm 0mm, clip, width=2.5in]{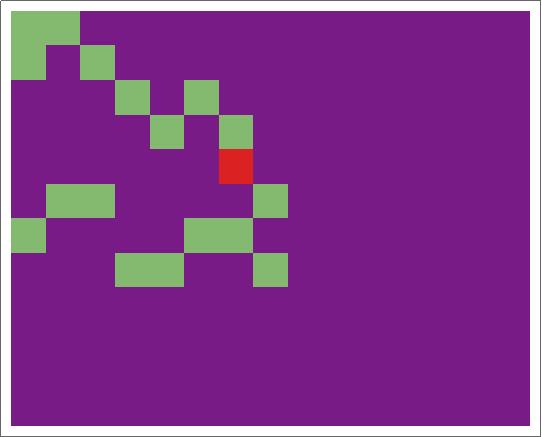}
&
(b)
\includegraphics[trim=0mm 0mm 0mm 0mm, clip, width=2.5in]{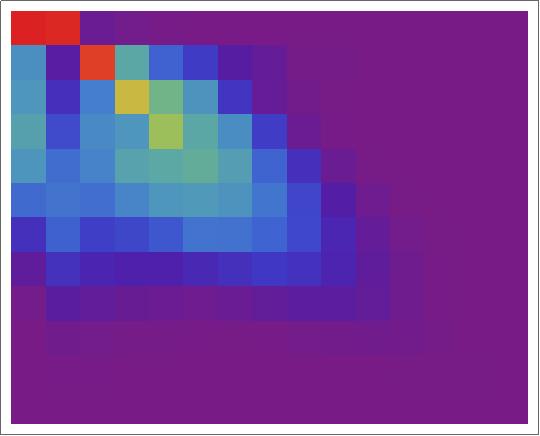}
\end{tabular}
}
\caption{{\sf {\small We realize the set of 7890 CICYs (Calabi-Yau threefolds as complete intersections  in products of projective spaces) as $12 \times 15$ matrices, padding with zeros where necessary. Then all CICY configurations are such matrices with entries in $\{0,1,2,3,4,5\}$.
We consider these as pixel colors and draw a typical CICY in (a), with 0 being purple.
In (b), we average over all such matrices component-wise, and draw the ``average'' CICY as a pixelated image.
}}
\label{f:cicy}
}
\end{figure}

Therefore, we can represent all the 7890 CICYs into $12 \times 15$ matrices over $\IZ/6\IZ$, embedded starting from the upper-left corner, say, and padding with zeros everywhere else.
A typical configuration (say, number 2000 in the list) thus becomes an image using Mathematica's \verb|ArrayPlot[ ]|
which is shown in part (a) of Figure \ref{f:cicy}.
Here, the purple background are the zeros, the greens are the ones and the one red block is 2.
Indeed, the CICY configuration matrices are rather sparse, dominated by 0 and following that, by 1.
In part (b) of the figure, we try to show the ``average'' CICY by component-wise average over all the $12 \times 15$ matrices.
Again, we see the largely purple (entry 0) background, with various shades mixed by the averaging process.
It is rather aesthetically pleasing to see the average CICY as an image.

It should be emphasized that such a pixelated depiction of CICY is {\it only} for visualization purposes and nothing more.
Usually, with graphic images, the best NN to use is so-called {\it convolution} layers where each pixel is convolved with its neighbours.
Our MLP does not have such a convolution layer so strictly speaking we are not taking advantage of the image-processing aspects of our data.
Nevertheless, this representation is very visual and one can check that {\it not} representing the CICY in matrix form and simply flattening the entire matrix into a long vector also achieves the same nice results \footnote{
I would like to thank Wati Taylor and Sanjaye Arora for many conversations and their helpful comments at \cite{conf2}, especially to Sanjaye for checking that this vector representation for CICYs also works well.
}.
Throughout the rest of this writing we will likewise use pixelation as a good visual guide for our geometries but will not be using any convolutional neural networks (CNN).

Now, our input space is much larger than that of the $W\IP^4$ case consider above:
$7890 \times 12 \times 15$ is 2 orders of magnitude larger than $7555 \times  5$, thus let us indulge ourselves with a full, rather than binary query.
That is, can we deep-learn, say the full list of Hodge numbers?
As usual, the Euler number is relatively easy to obtain and there is a combinatorial formula in terms of the integers $q_j^r$, whilst the individual Hodge numbers $(h^{1,1}, h^{2,1})$ involve some non-trivial adjunction and sequence-chasing, which luckily had been performed for us and occupied a sizeable portion of the exposition in the Bestiary \cite{cicy}.
Our warm-up exercise is to machine-learn this result.

Again, we set up a list of training rules (padded configuration matrix $\to h^{1,1}$), a typical entry, adhering to our example in Part (a) of Figure \ref{f:cicy}, would be
\begin{equation}
\left(
\begin{array}{c|c}
{\tiny
\begin{array}{cccccccc}
 1 & 1 & 0 & 0 & 0 & 0 & 0 & 0 \\
 1 & 0 & 1 & 0 & 0 & 0 & 0 & 0 \\
 0 & 0 & 0 & 1 & 0 & 1 & 0 & 0 \\
 0 & 0 & 0 & 0 & 1 & 0 & 1 & 0 \\
 0 & 0 & 0 & 0 & 0 & 0 & 2 & 0 \\
 0 & 1 & 1 & 0 & 0 & 0 & 0 & 1 \\
 1 & 0 & 0 & 0 & 0 & 1 & 1 & 0 \\
 0 & 0 & 0 & 1 & 1 & 0 & 0 & 1 \\
\end{array}
}
&
0  \\ \hline
0 & 0
\end{array}
\right)_{12 \times 15}
\longrightarrow
\qquad 
6 \ ;
\end{equation}
this is a CY3 of Hodge numbers $(h^{1,1},h^{2,1}) = (8,22)$ and hence Euler characteristic $\chi = -28$.
This training list we shall call \verb|dd|.

We now set up the neural network as
\begin{equation}\label{nnCY}
\begin{array}{l}
\verb|net = NetChain[{LinearLayer[1000], ElementwiseLayer[LogisticSigmoid],| \\
\verb|               LinearLayer[100], ElementwiseLayer[Tanh], SummationLayer[]},| \\
\verb|              "Input" -> {12, 15}];|
\end{array}
\end{equation}
We specified the input as \verb|{12,15}| and have allowed a quite large linear layer of 1000 weights at the input level to allow for parameters.
Next, we train the network with the list \verb|dd|, allowing for 1000 iterations to reduce error:
\begin{verbatim}
net = NetTrain[net, dd, MaxTrainingRounds -> 1000];
\end{verbatim}
Despite these sumptuous choice of parameters, the training takes about a mere 8 minutes, associating an optimized real number to each configuration (image) and a simple check of the round-up against actual $h^{1,1}$
\begin{verbatim}
Select[Table[{Round[net[dd[[i, 1]]]], dd[[i, 2]]}, {i, 1, 7890}],
         #[[1]] != #[[2]] &] // Length
\end{verbatim}
 returns 7, meaning that our network has been trained to an accuracy of $(7890-7)/7890 \simeq 99.91\%$ in under 10 minutes.
Thus, this is very much a ``learnable'' problem and the network functions well as a {\em classifier}, almost completely learning the Hodge data for CICYs.

\paragraph{Learning Curve: }
What about the network as a predictor, which is obviously a more salient question?
Let us repeat the methodology above, viz., to train with a portion of the input data, and see whether it could extrapolate to the full dataset.
Suppose the neural network sees only a random sample of 4000 of \verb|dd|, which is particularly pertinent were the classification not complete and which, though not here, would be the case with most problems in the business.

Using the above neural network takes about 6 minutes.
Then, checking against the full dataset comprising of configurations/images the MLP has never before seen, we find 184 discrepancies.
Considering 
(1) that we have only trained the network for a mere 6 minutes,
(2) that it has seen less than half of the data, 
(3) that it is a rather elementary MLP with only 5 forward layers, and 
(4) that the variation of the output is integral ranging from 0 to 19, with no room for continuous tuning, 
achieving $97.7\%$ accuracy and cosine distance 0.99 with so little effort is quite amazing!

\begin{figure}[!h!t!b]
\centerline{
\includegraphics[trim=0mm 0mm 0mm 0mm, clip, width=6in]{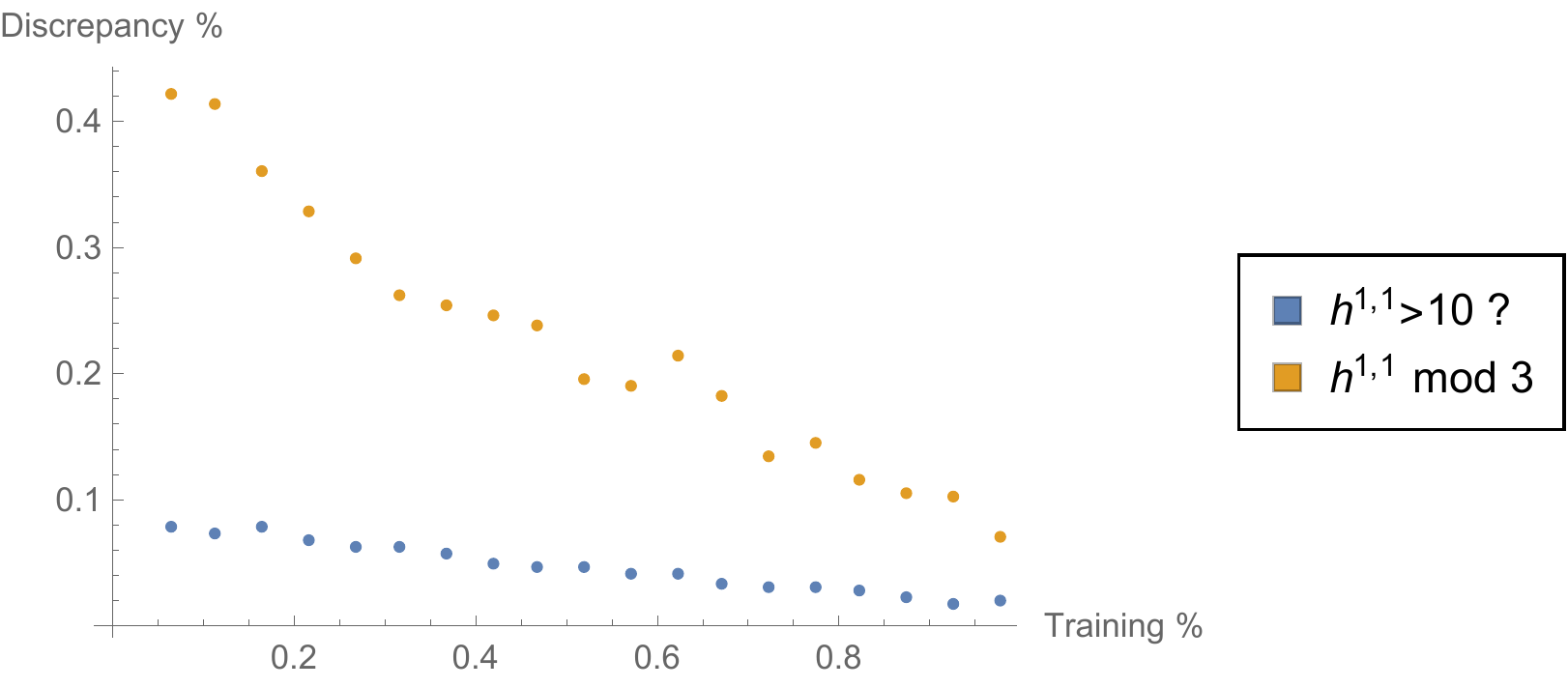}
}
\caption{{\sf {\small 
The learning curve for (a) large Kahler structure (whether $h^{1,1} > 10$) and (b) divisibility of K\"ahler parametres modulo 3, for the CICY manifolds.
We take, in increments of 400, and starting from 500 random samples, with which we train the NN (shown as percentage of the total data size of 7890 in the horizontal).
We then check against the full data to see the percentage discrepancy, shown in the vertical.
}}
\label{f:CICYlearningC}
}
\end{figure}

It is expedient to investigate the learning curve here, which is presented in Figure \ref{f:CICYlearningC}.
The training data is taken as random samples, in increments of 400, and starting from 500 from the total 7890, shown as a percentage in the horizontal axis.
We then validate against the total dataset of 7890 and see the percentage, drawn in the vertical, of discrepancies.
We see that the question of whether  $h^{1,1} > 10$ is extremely well-behaved: where 10\% discrepancy (90\% accuracy) is achieved with only 10\% training data, decreasingly steadily with increasing training data.
Such a trend is even more dramatic for testing whether $h^{1,1}$ is divisible by 3: more than 40\% discrepancy at 10\% available data, decreasing to less than 10\% at full data.
Nevertheless, it is clear that testing whether a quantity exceeds a given value is better than testing its divisibility properties.

The general strategy when confronted with a difficult computation, especially in algebraic geometry, is therefore clear.
Suppose we have a large or even unknown classification and we have handle only for a fraction of requisite geometrical/topological quantities.
We can select, starting from a small percentage of known results, establish a NN, and see the trend of the learning curve.
If the discrepancies are obviously decreasing in a satisfactory manner, then we gain further confidence in predicting what the required quantity is over the entire dataset.

\subsubsection{Four-folds}
The generalization from CICY 3-folds to 4-folds, with \eqref{cicy} altered mutatis mutandis (e.g., making $K = \sum\limits_{r=1}^m n_r-4$), has more recently been accomplished \cite{Gray:2013mja} and the dataset is, understandably, much larger.
Here, we have 921,497 configuration matrices of size $1 \times 1$  (corresponding to the sextic $[5|6]$) up to 16 rows and 20 columns, taking values in 0 to 6, with Euler characteristic ranging from 0 to 2610.
The computation took an impressive 7487 CPU hours, almost 1 entire year if on a single core.

As before, we can consider each CICY4 as an $16 \times 20$ pixelated image, padding with zeros where necessary, with 7 shades of colour.
We thus have about 1 million images to learn from.
\begin{figure}[t]
\centerline{
\begin{tabular}{cc}
(a)
\includegraphics[trim=0mm 0mm 0mm 0mm, clip, width=2in]{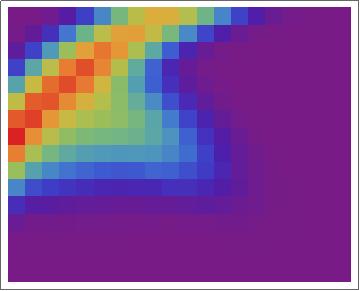}
&
(b)
\includegraphics[trim=0mm 0mm 0mm 0mm, clip, width=3in]{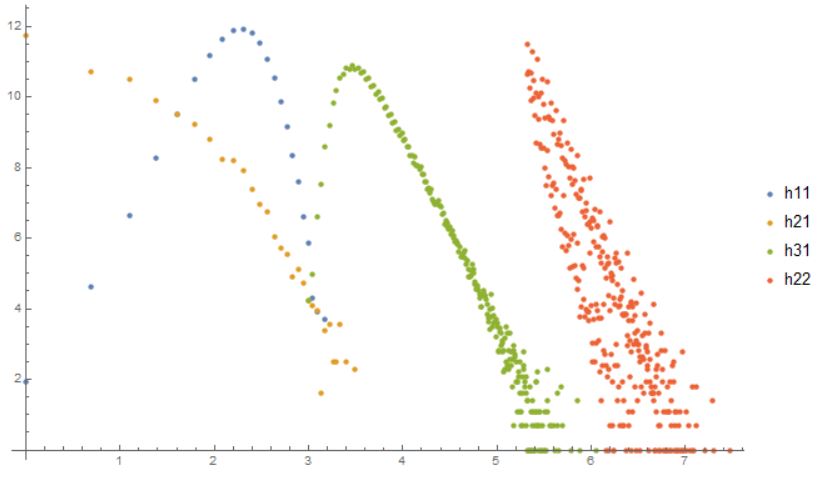}
\end{tabular}
}
\caption{{\sf {\small We realize the set of 921,497 CICY4s  as $16 \times 20$ matrices (bottom-right padding with zeros where necessary) with entries in $\{0,1,2,3,4,5,6\}$.
Each is then a pixelated images with 7 shades of colour with 0 being, say, purple.
In (a), we average over all such matrices component-wise, and draw the ``average'' CICY4, in contrast to part (a) of Figure \ref{f:cicy}.
For reference, we include, in part (b), the histogram of the (now four) Hodge numbers: $(h^{1,1}, h^{2,1}, h^{3,1}, h^{2,2})$; because of the spread in values, we perform a log-log plot.
}}
\label{f:cicy4}
}
\end{figure}
As above, the Euler characteristic has a simple combinatorial formula in terms of the $q_j^r$ and it is also an alternating sum in the 4 non-trivial Hodge numbers $(h^{1,1}, h^{2,1}, h^{3,1}, h^{2,2})$.
In Figure \ref{f:cicy4}, we plot the average over all the images by component-wise arithmetic mean in part (a) and in part(b), for reference, we show the distribution of the four Hodge numbers on a log-log plot.
Let us try to deep-learn these Hodge numbers. 
With the same network as in \eqref{nnCY}, but with \verb|"Input" -> {16,20}|, we purposefully take {\it incomplete} knowledge by using only the first 20,000 data points as training data.
For $h^{1,1}$, which ranges between 1 to 24 in the training set, we achieve almost complete accuracy (22 discrepancies) within about an hour of deep-learning.

Encouraged, we try to extrapolate: testing against the first 30,000 of the full dataset, which takes about 30 minutes, we find 23,521 matches, meaning that we have achieved $78.4\%$ accuracy within 2 hours, much less than the time to compute $h^{1,1}$ from first principles.
Similarly, for $h^{2,1}$, which ranges from 0 to 33, we achieve similar figures.
The remaining Hodge numbers, $h^{3,1}$ and $h^{2,2}$ have a much wider spread, taking values in $[21,426]$ and $[204,1752]$ respectively, would require more than 20,000 to train to greater confidence.

There are many more properties of CICYs, and indeed of CY manifolds in general, comprising an ongoing industry, from identifying torus or K3 fibration structure for F-theory to adding orientifolds, which we could deep-learn.
For now, let us move onto a closely related subject which has over the last 2 decades vastly generalized the computation of Hodge numbers.

\subsection{Bundle Cohomology}
The subject of vector bundle cohomology has, since the so-named ``generalized embedding'' \cite{Candelas:1985en} of compactifying the heterotic string on smooth Calabi-Yau threefolds $X$ endowed with a (poly-)stable holomorphic vector bundle $V$, become one of the most active dialogues between algebraic geometry and theoretical physics.
The realization \cite{Braun:2005nv} that the theoretical possibility of \cite{Candelas:1985en} can be concretely achieved by a judicious choice of ($X,V)$ to give the exact MSSM spectrum induced much activity in establishing relatively large datasets to see how often this might occur statistically \cite{He:2009wi,Anderson:2007nc,He:2013epn,Gabella:2008id,Candelas:2008wb}, culminating in \cite{Anderson:2013xka,Anderson:2012yf} which found some 200 out of a scan of $10^{10}$ bundles which have exact MSSM content.

Upon this vast landscape let us take an insightful glimpse by taking the dataset of \cite{Gabella:2008id}, which are $SU(n)$ vector bundles $V$ on elliptically fibred CY3.
By virtue of a spectral-cover construction \cite{ellip,Friedman:1997yq}, these bundles are guaranteed to be stable and hence preserves $\cN=1$ supersymmetry in the low-effective action, together with GUT gauge groups $E_6$, $SO(10)$ and $SU(5)$ respectively for $n=3,4,5$.

We take the  base of the elliptic fibration - of which there is a finite list \cite{Morrison:1996na} - as the $r$-th Hirzebruch surface ($r = 0, 1, \ldots, 12$ denoting the inequivalent ways which $\IP^1$ can itself fibre over $\IP^1$ to give a complex surface), in which case the stable $SU(n)$ bundle is described by 5 numbers $(r,n, a,b, \lambda)$, with $(a,b) \in \IZ_+$ and $\lambda \in \IZ/2$ being coefficients which specify the bundle via the spectral cover.
This ordered 5-vector will constitute our neural input.
The database of viable models were set up in \cite{Gabella:2008id}, viable meaning that the bundle-cohomology groups of $V$ are such that
\begin{equation}
h^0(X,V) = h^3(X,V) = 0 \ , \quad
\left| h^1(X,V) - h^2(X,V) \right| \equiv 0 \ (\bmod 3) \ ,
\end{equation}
where the first is a necessary condition for stability and the second, that the GUT theory has the potential to allow for 3 net generations of particles upon breaking to MSSM by Wilson lines.
Over all the Hirzebruch-based CY3, 14,264 models were found; a sizeable play-ground.

Suppose we wish the output to be a 2-vector, indicating (I) what the gauge group is, as denoted by $n$, and (II) whether there are more generations than anti-generations, as denoted by the sign of the difference $h^1(X,V) - h^2(X,V)$; this is clearly a phenomenologically interesting question.
The fact that the network needs to produce a vector output need not worry us; we replace the last summation layer of \eqref{nnCY} by a linear layer of matrix size 2:
\begin{verbatim}
net = NetChain[{LinearLayer[1000], ElementwiseLayer[LogisticSigmoid], 
            LinearLayer[100], ElementwiseLayer[Tanh], LinearLayer[2]},
       "Input" -> {5}];
\end{verbatim}
With 1000 training rounds as before, and with the dataset consisting of replacements of the form
\[
(r,n, a,b, \lambda) \to (n, {\rm Sign}( h^1(X,V) - h^2(X,V)))
\] 
(note that we purposefully kept $n$ common to both input and output as a cross-check and to see how the network responds to vector output), in about 10 minutes (the vector output and thus the last linear layer slows the network by a factor of 2 compared to previous trainings), we achieve 100\% accuracy (i.e., the neural network has completely learnt the data).
This seems like an extremely learnable problem.
Indeed, training with partial data, say 100 points, less than 1\% of the total data, achieves 99.9\% predicative accuracy over the entire set!

Similarly, we could query whether $h^1(X,V) - h^2(X,V)$ is even or odd, which is also an important question because it dictates the parity of the size of the fundamental group (as a finite discrete symmetry group) of the CY3, which is a difficult issue to settle theoretically.
Using the above network, a 67\% accuracy can be attained in 10 minutes training on 8000 points
This shows that the pattern here is indeed more difficult to recognize as compared to the sign, as is consistent with expectations.
There are endless variations to this type of questions and deep-learning and we shall discuss some pertinent ones in the prospectus; 
for now, let us match onward to further available data.

\subsection{Aspects of the KS Dataset}
The largest dataset in Calabi-Yau geometry, and indeed to our knowledge in pure mathematics, more than such online resources as modular forms or dessins d'enfants, is the Kreuzer-Skarke list of reflexive (convex lattice) polytopes \cite{ks}.
There are 4319 in dimension 3 and 473,800,776 in dimension 4.
Much work, still ongoing, has been focused on the latter because the anticanonical divisor within the associated toric variety defines a smooth Calabi-Yau threefold \cite{cyDB,cyDBus}.
We will leave explorations and deep-learning of this formidable set to future work.
For now, let us focus on the 4319 reflexive lattice polyhedra which has recently been harnessed for certain Sasaki-Einstein volume conjectures \cite{He:2017gam}.

First, we can represent the data in a conducive way.
Each polyhedron $\Delta$ consists of a list of integer vertices as 3-vectors (of which $\Delta$ is the convex hull); there is a single interior lattice point $(0,0,0)$ and we can record all the $n-1$ lattice points on the boundary facets.
Hence, each $\Delta$ is a $3 \times n$ integer matrix $M$, with one column being $(0,0,0)$.
It turns out that $n$ ranges from 5 to 39 (the 5 clearly including the lattice tetrahedron enclosing the origin), and the entries of $M$ take values in
$\{-21, -20, \ldots, 5,6\}$.

\begin{figure}[!h!t!b]
\begin{tabular}{cl}
(a)
&
\includegraphics[trim=0mm 0mm 0mm 0mm, clip, width=5in]{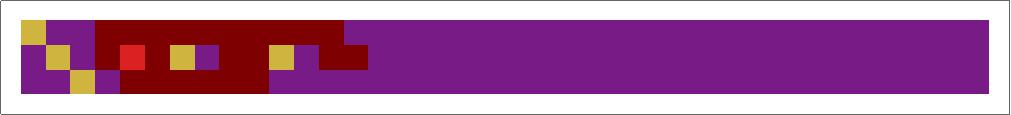}
\\
(b)
&
\includegraphics[trim=0mm 0mm 0mm 0mm, clip, width=5in]{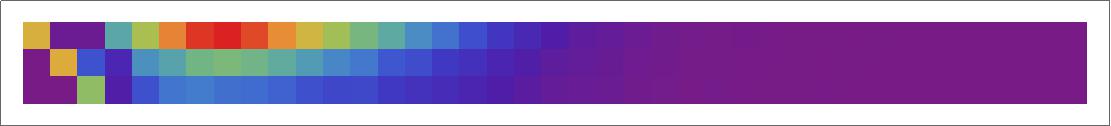}
\end{tabular}
\caption{{\sf {\small We realize the set of 4319 reflexive polyhedra as $3 \times 39$ matrices 
(right padding with zeros where necessary) with entries in $\{-21, -20, \ldots, 5,6\}$.
Each is then a pixelated image with 28 shades of colour with 0 being, say, purple.
In (a), we show a typical sample;
in part (b), we show the average by component-wise root mean square. 
Again, the image is purely for amusement; we are not using a CNN in this paper.
}}
\label{f:KS3}
}
\end{figure}

Hence, each $\Delta$ is a $3 \times 39$ pixelated image (right padding where necessary), with 28 shades of colour.
We show a random sample  (number 2000 in KS dataset)   in parts (a) of Figure \ref{f:KS3}.
In part (b), because we now have a mixture of positive and negative entries to the matrix, we compute the square-root of the component-wise 
sum of the squares to the 4319 configurations, giving an idea of the ``average'' appearance of the reflexive polyhedra. 

To each $\Delta$ we can associate two different geometries:
\begin{itemize}

\item a compact smooth Calabi-Yau 2-fold, or K3-surface $\cK$, being a hypersurface as the anti-canonical divisor in the Fano toric variety $X(\Delta)$ constructed from the polytope.
These are algebraic K3 surfaces so the Hodge numbers are all the same. The Picard number $Pic(\cK)$, one the other hand, is a non-trivial property and has been computed \cite{ks}.
The reader is referred to Fig.~16 of \cite{He:2015fif} for a distribution of these Picard numbers;

\item a non-compact affine Calabi-Yau 4-fold which is the total space of the anti-canonical bundle over $X(\Delta)$; this Calabi-Yau space is itself a real cone over a compact Sasaki-Einstein 7-manifold $\cY$.
The various Z-minimized volumes $V(\cY)$ of $\cY$, normalized with respect to $S^7$, are algebraic numbers computed in \cite{He:2017gam}
(q.v.~ibid., Fig.~10 for a distribution of these volumes);
\end{itemize}

We will now deep-learn $Pic(\cK)$ and $V(\cY)$ with respect to our images.
The data size here is relatively small compared to the previous so once again we will use a binary function for the volume, say, if $V(\cY) > 200$ then 1, and 0 otherwise. Using the same simple network as \eqref{cicy}, but with input size $3 \times 39$, we achieve $57\%$ accuracy in about 5 mins.
Similarly, for the Picard numbers, which range from 0 to 19, the direct learning can achieve $53\%$ accuracy in equal time, and due to the discrete nature of the output, had we chosen a step function, say 1 if $Pic(\cY) > 10$ and 0 otherwise, then a $59\%$ accuracy is attained.
Of course, these figures are not as impressive as the previous case studies, but this is precisely due to the relative paucity of data: less than 5000.
As one would expect, deep-learning must be accompanied by ``big data''.

A plethora of such big data we certainly have.
The 473 million reflexive polytopes in dimension 4 is the ne plus ultra of our landscape data.
In fact, because of the different possibilities of triangulations, the number of Calabi-Yau threefolds obtained therefrom is expected to be many orders of magnitude larger than this and is still very much work in progress.
To deep-learn this data is ipso facto a significant endeavour which we shall leave for a major future undertaking.

For now, let us take a sample of 10,000 from the peak-distribution \cite{He:2015fif} of Euler characteristic zero, i.e., self-mirror, manifolds.
Here, the configuration can be padded into $4 \times 21$ integer matrices taking integer values in $[-14,9]$ and the individual Hodge numbers $h^{1,1} = h^{2,1} \in \{14,15,16,17,18\}$.
This is therefore a well-suited classifier problem, with input pixelation of $4 \times 21$ with 24 shade and 5-channel output.
Again, we train with partial data, say the first 4000, and then test against the full 10,000 to see its predictive power.
In under 5 minutes, the neural network (with modified input dimension) in \eqref{nnCY}, gives 61\% accuracy in predicting the full-set and 78\% in fitting the 4000. 

\subsection{Quiver Gauge Theories}
As a parting example, having been emboldened by our venture, let us tackle affine varieties in the context of quiver representations.
Physically, these correspond to world-volume gauge theories coming from D-brane probes of geometric singularities in string theory, as well as as the space of vacua for classes of supersymmetric gauge theories in various dimension; they have been data-mined since the early days of AdS/CFT (cf.~\cite{Hanany:1998sd,Feng:2000mi}).
When the geometry concerned is an affine toric Calabi-Yau variety, the realization of brane-tiling \cite{Franco:2005sm} has become the correct way to understand the gauge theory and since then databases have begun to be compiled \cite{Davey:2009bp,Franco:2017jeo}.

As far as the input data is concerned, however, it is straight-forward; it consists of a quiver (as a directed graph, with label 1 as dimension vector for simplicity) and a relation on the quiver imposed by the Jacobian of a polynomial super-potential (q.v.~\cite{He:2013epn} for a rapid review).
For example, the following is a typical quiver:
\begin{equation}
\begin{array}{cc}
\begin{array}{c}\includegraphics[trim=0mm 0mm 0mm 0mm, clip, width=2in]{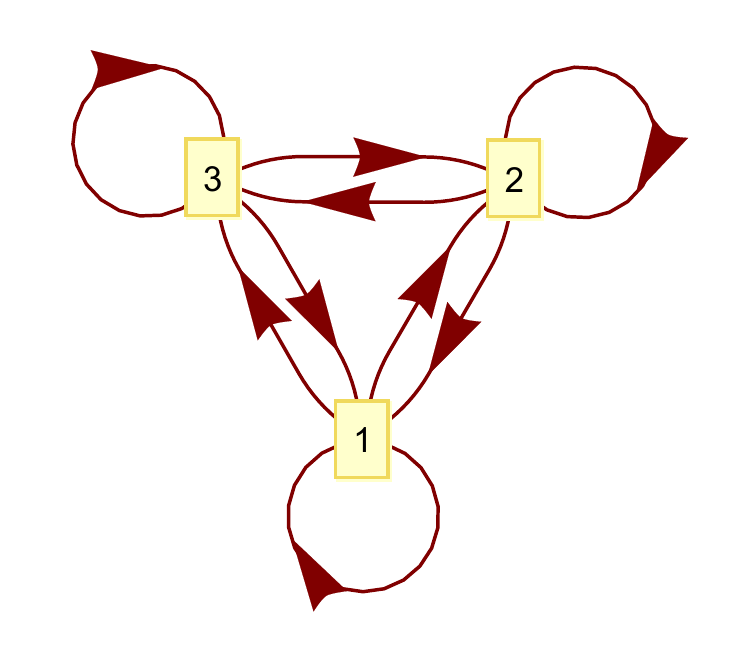}\end{array}
&
\begin{array}{rcl}
W & = & X_{12}.\phi _2.X_{21}-X_{21}.\phi _1.X_{12}+X_{23}.\phi _3.X_{32}-\\
	&& -X_{32}.\phi _2.X_{23}+\phi _1.X_{13}.X_{31}-\phi_3.X_{31}.X_{13} \ .
\end{array}
\end{array}
\end{equation}
In the above, there are 3 nodes in the quiver and we have denoted arrows $a \to b$ as $X_{ab}$ if $a \neq b$ and as $\phi_a$ if it is an arrow joining node $a$ to itself (for $a,b=1,2,3$).
There are thus a total of 9 arrows, which can be interpreted as space-times fields in the gauge theory.
The superpotential $W$ is here a cubic polynomial in the 12 arrows, whose Jacobian $\partial W$ imposes the set of relations.
The representation variety is given as the GIT quotient of $\partial W$ by the gauge-fixing conditions by the Eulerian cycles in the quiver, and is here the affine variety
$\IC \times \IC^2 / (\IZ/3\IZ)$ which is a local calabi-Yau threefold, being a direct product of $\IC$ and an orbifold surface singularity of type A.

We can succinctly encode the above information into two matrices:
\begin{enumerate}
\item D-term matrix $Q_D$, which comes from the kernel of the incidence matrix $d$ of the quiver, a $3 \times 12$ matrix each column of which corresponds to an arrow with $-1$ as head and $+1$ as tail and 0 otherwise;
\item F-term matrix $Q_F$, a $2 \times 12$ matrix each column of which documents where and with what exponent the field corresponding to the arrow appears in $\partial W$.
\end{enumerate}

Concatenating $Q_D$ and $Q_F$ gives the so-call total charge matrix $Q_t$ of the moduli space as a toric variety (q.v.~\S2 of \cite{Feng:2000mi} for the precise procedure).
For the above example, the incidence matrix $d$, the Jacobian of the superpotential, and thence the total charge $Q_t$ matrices are
\begin{equation}
\begin{array}{l}
d =
{\scriptsize
\left(
\begin{array}{ccccccccc}
 0 & -1 & -1 & 1 & 0 & 0 & 1 & 0 & 0 \\
 0 & 1 & 0 & -1 & 0 & -1 & 0 & 1 & 0 \\
 0 & 0 & 1 & 0 & 0 & 1 & -1 & -1 & 0 \\
\end{array}
\right)
}
\\
\partial W = 
\left\{
\begin{array}{l}
x_3 x_7-x_2 x_4, \ x_4 x_5-x_1 x_4, \\
x_2 x_5-x_1 x_2, \ x_2 x_4-x_6 x_8, \\ 
x_1 x_7-x_7 x_9, \ x_1x_3-x_3 x_9, \\
x_8 x_9-x_5 x_8, \ x_6 x_9-x_5 x_6, \\
x_6 x_8-x_3 x_7
\end{array}
\right.
\end{array}
\leadsto
Q_t = 
{\scriptsize
\left(
\begin{array}{c|c}
Q_D &
\begin{array}{ccccccccc}
 4 & -1 & -2 & 0 & -1 & 0 & 0 & 0 & 0 \\
 2 & -2 & -1 & 0 & 1 & 0 & 0 & 0 & 0 \\
\end{array}
\\
\hline
Q_F &
\begin{array}{ccccccccc}
 2 & -1 & -1 & 0 & -1 & 0 & 0 & 1 & 0 \\
 1 & 0 & -1 & 0 & -1 & 0 & 1 & 0 & 0 \\
 1 & -1 & 0 & 0 & -1 & 1 & 0 & 0 & 0 \\
 1 & -1 & -1 & 1 & 0 & 0 & 0 & 0 & 0 \\
\end{array}
\end{array}
\right)
}
\end{equation}
where we have indexed the 9 arrows, in accord with the columns of $d$, as $x_{i=1,\ldots,9}$.

The combinatorics and geometry of the above is a long story spanning a lustrum of research to uncover followed by a decade of still-ongoing investigations.
However, for our present purposes, we will only consider the theories to be data for the neural network to learn.
In the first database of \cite{Davey:2009bp}, a host of examples (albeit having many inconsistent theories as D-brane world-volume theories) were tabulated.
A total of 375 quiver theories much like the above were catalogued (a catalogue which has recently been vastly expanded in \cite{Franco:2017jeo}, which also took care of ensuring physicality).
Though not very large, this gives us a playground to test some of our ideas.
The input data is the total charge matrix $Q_t$, the maximal of whose number of rows and columns are, respectively 33 and 36 , and all 
taking values in $\{-3,-2, \ldots, 3,4\}$.

\begin{figure}[!h!t!b]
\centerline{
\includegraphics[trim=15mm 0mm 0mm 0mm, clip, width=6in]{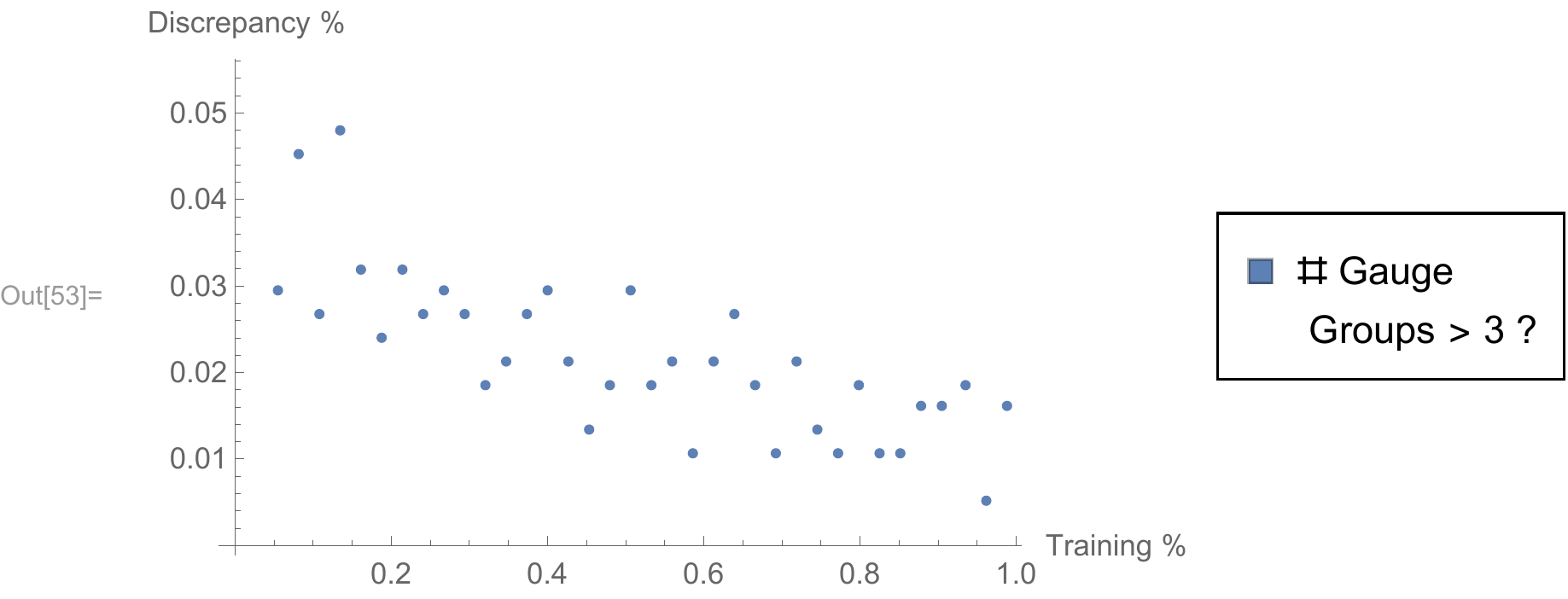}
}
\caption{{\sf {\small 
The learning curve for the number of gauge groups for brane tilings.
We take, in increments of 10, and starting from 10 random tilings, with which we train the NN (shown as percentage of the total data size of 375 in the horizontal).
We then check against the full data to see the percentage discrepancy, shown in the vertical.
}}
\label{f:dimertrainingC}
}
\end{figure}

Now, suppose we wish to know the number of points of the toric diagram associated to the moduli space, which is clearly an importantly quantity.
In principle, this can be computed: the integer kernel of $Q_t$ should give a matrix whose columns are the coordinates of the toric diagram, with multiplicity.
Such multiplicity had been realized, via the brane-tiling story, to be associated to the perfect matchings of the bipartite tiling.
Nonetheless, one could train the network without any knowledge of this (where the functions \verb|pad[ ]| and the list \verb|Tilings| are self-explanatory):
\begin{verbatim}
data = Table[pad[("Qt" /. Tilings[[i]])] -> 
                    Length[Union["Points" /. ("Diag" /. Tilings[[i]])]], 
       {i, 1, Length[Tilings]}];
ClearAll[net];
net = NetChain[{LinearLayer[1000], ElementwiseLayer[LogisticSigmoid],
               LinearLayer[100], ElementwiseLayer[Tanh], SummationLayer[]},
               "Input" -> {33, 36}];
net = NetTrain[net, data, MaxTrainingRounds -> 1000];
\end{verbatim}
Training with our network with the full list achieves, in under 5 minutes, $99.5\%$ accuracy. 
We now plot the learning curve, as shown in Figure \ref{f:dimertrainingC}. We see that the discrepancy decreases from 5\% to less than 1\% as we increase the percentage training data (taken as a random sample), validating against the full data.
Thus, this problem is one perfectly adapted for the NN, with high accuracy acheived even for small training sample.

\subsubsection{A Hypothetical Input}
Now, let us attempt something more drastic.
The step to go from $d$ to $Q_d$ is expensive and in fact constituted the major hurdle to understanding D-brane gauge theories before tiling/dimer-technology;
it involves finding Hilbert basis of lattice points and dual cones in very high dimensions.
Even with the latest understanding and implementation of tilings, finding perfect matchings is still non-trivial \cite{Franco:2017jeo}.
Yet, we know that the input data $d$ and $Q_F$, which can be written down rapidly from definition, {\it must} at some level determine the output.
This is thus a perfect problem adapted to  machine learning.

Let us create a matrix, of size $41 \times 52$, which will accommodate the stacking of $d$ onto $Q_F$, right padding with 0 as necessary.
It should again be emphasized that computationally, this matrix has no meaning with regard to the toric algorithms developed over the years.
The entries of the matrix take integer values in $[-2,3]$ and for us, this is formally a $41 \times 52$ pixelated image with 6 shades of colour.
Thus a typical data would be of the form (say, for the $\IC \times \IC^2 /(\IZ/ 3\IZ)$ example above),
\begin{equation}
\begin{array}{c}\includegraphics[trim=0mm 0mm 0mm 0mm, clip, width=1.5in]{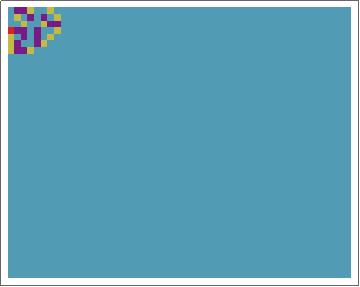}\end{array}
\qquad
\longrightarrow
\qquad
6 \ ,
\end{equation}
with magenta here denoting 0.
Training with the above network still achieves $98.0\%$ accuracy in learning the entire dataset in less than 5 minutes.
Even with 200 random samples, the network can predict to an accuracy of $66.4\%$ in minutes, 
which is expected to increase with training time as well as sample size.

\subsection{A Reprobate}
Lest the readers' optimism be elevated to unreasonable heights by the string of successes with the neural networks, it is imperative that we be aware of deep-learning's limitations.
We therefore finish with a sanity check that a neural network is not some omnipotent magical device, nor an oracle capable of predicting {\it any} pattern.
An example which must be doomed to failure is the primes (or, for that matter, the zeros of the Riemann zeta function).
Indeed, if a neural network could learn some unexpected pattern in the primes, this would be a rather frightening prospect for mathematics.

Let us thus take our reprobate example to be the following simple dataset
\begin{verbatim}
   dd = Table[ i -> Prime[i], {i, 1, 5000} ];
\end{verbatim}
Training with our neural network \eqref{nnCY} and counting discrepancies as always
\begin{verbatim}
Select[Table[Round[net[dd[[i, 1]]]]-dd[[i, 2]], {i, 1, 5000}],# == 0 &]//Length
\end{verbatim}
gives 5 after some 10 minutes.
That is, of the first 5000 primes, the network has learnt only 5.

Perhaps having more sophisticated input would help? We can attempt with a data-structure like
\begin{verbatim}
dd = Table[   Table[If[j <= i, Prime[j], 0], {j, 1, 5000}] -> Prime[i + 1], 
        {i, 1, 5000}];
\end{verbatim}
so that the input is now a zero-right-padded vector associating the $n$-th prime to the list of all its precedents.
Again, we still achieve no better than a 0.1\% accuracy.

Our neural network is utterly useless against this formidable challenge; we are better off trying a simple regression against some $n \log(n)$ curve, as dictated by the prime number theorem.
This is a sobering exercise as well as a further justification of the various case studied above, that it is indeed meaningful to deep-learn the landscape data and that our visual representation of geometrical configurations is an efficient methodology.

It is also interesting to find, through our experience, that NNs are good with algebraic geometry but not so much with number theory.
This principle could be approximately understood.
At the most basic level, every computation in algebraic geometry, be it a spectral sequence or a Gr\"obner basis, reduces to finding kernels and cokernels of sets of matrices (over $\IZ$ or even over $\IC$), albeit of quickly forbidding dimensions. Matrix/tensor manipulation is the heart of any NN.
Number theory, on the other hand, ultimately involves patterns of prime numbers which, as is well known, remain elusive.

\section{Conclusions and Prospectus}\label{s:conc}
There are many questions in theoretical physics, or even in pure mathematics, for which one would only desire a qualitative, approximate, or partial answer, and whose full solution would often either be beyond the current scope, conceptual or computational, or would have taken considerable effort to attain.
Typical such questions could be ``what is the likelihood of finding a universe with three generations of particles within the landscape of string vacua or inflationary scenarios'', or ``what percentage of elliptic curves have L-functions with prescribed poles''?
Attempting to address these profound questions have, with the ever-increasing power of computers, engendered our community's version of ``big data'', which though perhaps humble compared to some other fields, do comprise, especially considering the abstract nature of the problems at hand, of significant information often resulting from intense dialogue between teams of physicists and mathematicians for many years.

On the still-ripening fruits of this labour the philosophy of the last decade or so, particularly for the string phenomenology and computational geometry community, has been to (I) create larger and larger datasets and (II)  scan through them to test the likelihood of certain salient features.
Now that the data is augmenting in size and availability, it is only natural to follow the standard procedures of the data-mining community.
In this paper, we have proposed the paradigm of applying deep-learning, via neural networks, such data.
The purpose is twofold, the neural network can act as
\begin{description}
\item[Classifiers ] by association of input configuration with a requisite quantity, and pattern-match over a given dataset;
\item[Predictors ] by extrapolating to hithertofore unencountered configurations, having deep-learnt a given (partial) dataset.  
\end{description}
This is, of course, the archetypal means by which Google deep-learns the internet and hand-writing recognition software adapts to the reader's esoteric script.
Indeed, this is why {\it neural networks are far more sophisticated than usual logistic regression models} where a single analytic function with a set of given parameters is used to ``best-fit''. The network in general consists of a multitude of such functions, not necessarily analytic but rather algorithmic, which mutually interact and recursively optimize.

It is intriguing that by going through a wealth of concrete examples from what we have dubbed {\bf landscape data}, some of whose creation the author had been a part, this philosophy remains enlightening.
Specifically, we have taken test cases from a range of problems in mathematical physics, algebraic geometry and representation theory, such as Calabi-Yau datasets, classification of stable vector bundles, and catalogues of quiver varieties and brane tilings. 
We subsequently saw that even relatively simple neural networks like the multi-layer perceptron can deep-learning to extraordinary accuracy.

In some sense, this is not surprising, there is underlying structure to any classification problem in our context, which may not be manifest.
Indeed, what is novel is to look at the likes of a complete-intersection Calabi-Yau manifold or an integer polytope as a pixelated image, no different from a hand-written digit, for whose analysis machine-learning has become the de facto method and a blossoming industry.
\begin{quote}
{\it
The landscape data, be they work of human hands, elements of Nature or conceptions of Mathematics, have inherent structure, sometimes more efficiently uncovered by AI via deep-learning.
}
\end{quote}
Thereby, one can rapidly obtain results, before embarking on finding a reductionist framework for a fundamental theory explaining the results or proceed to intensive computations from first principles.
This paradigm is especially useful when classification problems become intractable, which is often the case, here a pragmatic approach would be to deep-learn partial classification results and predict future outcome (for this, the predictive accuracy becomes an important issue which we will address momentarily).

Under this rubric, the possibilities are endless.
Several immediate and pertinent directions spring to mind.
\begin{itemize}
\item In this paper we have really only used NNs are predictors in the sense of {\it supervised training}.
This addresses well the generic problems in, for instance, string phenomenology: we are given some classification of geometries through which we wish to sift in order to find a desired model (such as the MSSM); the classifications are large and we in principle know what we wish to compute though we have the computational power to perform only a fraction of the cases.
Therefore we train an NN in a supervised way, with a specific input and output in mind, in order to predict, with some confidence, the remaining intractable cases.
To use neural networks in an unsupervised way, where they find patterns as classifiers, would be an entirely new direction to explore.

\item The largest dataset in algebraic geometry/string theory is the Kreuzer-Skarke list \cite{ks,cyDB,cyDBus} of reflexive polytopes in dimension 4 from each of which many Calabi-Yau manifolds (compact and non-compact) can be constructed. 
  To discover hidden patterns is an ongoing enterprise \cite{He:2015fif,Taylor:2012dr} and the help of deep-learning would be a most welcome one.
  Moreover, it is not known how many compact CYs can arise from inequivalent triangulations of the polytopes and a systematic scan had only been done to small Hodge numbers \cite{cyDBus}; we could use the neural network's predictive power to extrapolate to the the total number of triangulations, and hence estimate the total number of Calabi-Yau threefolds. 

\item The issue of bundle stability and cohomology is a central problem in heterotic phenomenology as well as algebraic geometry.
  In many ways, this is a perfect problem for machine-learning: the input is usually encodable into an integer matrix or a list of matrices, representing the coefficients in an expansion into effective divisor classes, the output is simply a vector of integers (in the case of cohomology) or a binary answer (with respective to a given K\"ahler class, the bundle is either stable or not).

  The brute-force way involves the usual spectral sequences and determining all coboundary maps or finding the lattices of subsheafs, expensive by any standards. In the case of stability checking, this is an enormous effort to arrive at a yes/no query. With increasing number of explicitly known examples of stable bundles constructed from first principles, to deep-learn this and then estimate the probability of a given bundle being stable would be tremendous time-saver.

\item Whilst we could continue to list more prospective projects, there are theoretical issues which have arisen in all the above.
  First, there is a matter of convergence: how does increasing the complexity of the neural network or its training time decrease the error.
  There are detailed asymptotic studies of this \cite{HKP,Hassoun,Haykin} which should be taken into consideration.
  Second, every prediction must be accompanied by a confidence. The built-in \verb|Classify[ ]| command gives a probability for each extrapolated data point. For our neural networks, we have done a naive comparison to untrained data to get an idea but, especially when dealing the terra incognita of yet unclassified data, we need to specify and investigate the confidence level.

  For example, an ideal statement would be: here is a newly constructed manifold, we bearly know its properties and computing them from scratch would be difficult, however, based on similar manifolds classified before whose landscape we have deep-learnt, we can say with confidence $a$ that its Betti numbers are $b_1,b_2,\ldots$.
\end{itemize}

We hope the reader has been persuaded by not only the scope but also the feasibility of our proposed paradigm, a paradigm of increasing importance in an Age where even the most abstruse of mathematics or the most theoretical of physics cannot avoid compilations of and investigations on perpetually growing datasets.
The case studies of deep-learning such landscape of data here presented are but a few nuggets in an unfathomably vast gold-mine, rich with new science yet to be discovered.

\section*{Acknowledgments}
{\it Catharinae Sanctae Alexandriae adque Majorem Dei Gloriam.}\\
I am grateful to V.Jejjala, L.~Motl (and his kind words), B.~Nelson, G.~Sankaran, R.~Schimmrigk, R.-K.~Seong and R.~Thomas for 
comments on the first version of the paper, as well as to S.~Arora W.~Taylor for the many helpful conversations at the String Data conference 
\cite{conf2} which J.~Halverson, C.~Long and B.Nelson have brilliantly organized.

Indeed, I am indebted to the Science and Technology Facilities Council, UK, for grant ST/J00037X/1,
the Chinese Ministry of Education, for a ChangJiang Chair Professorship at NanKai University,
and the city of Tian-Jin for a Qian-Ren Award. Above all, I thank Merton College,
Oxford for continuing to provide a quiet corner of Paradise for musings and contemplations.

{\small

}

\end{document}